# Optical and Modal Features of Hemielliptic Dielectric Lenses

A.V. Boriskin[1,3], R. Sauleau[2], and A. I. Nosich[1,3]

[1] Institute of Radiophysics and Electronics NASU, Kharkov, Ukraine (a_boriskin@yahoo.com)
[2] Institut d'Electronique et de Télécommunications de Rennes, Université de Rennes 1, 35042 Rennes, France
[3] Université Européenne de Bretagne, Rennes, France

*Abstract*— Any dielectric lens has a finite closed boundary and therefore is, in fact, an open dielectric resonator capable of supporting resonant modes whose *Q*-factor depends of the lens parameters (size, shape, and material). The hemielliptic lens, that is an essential building block of many mm-wave and THz antennas, is not an exception: it supports the so-called half-bowtie (HBT) resonances that can strongly affect performance of such antennas. In this paper we illustrate the interplay between the optical and modal features in the electromagnetic behaviour of hemielliptic lenses and highlight the drastic influence of the HBT resonances on radiation characteristics of lens antennas. We also discuss the difficulties associated with accurate description of the resonant phenomena in compact-size hemielliptic lenses with conventional techniques and provide recommendations on how to minimize the parasitic impact of HBT resonances on the antenna performance.

## I. INTRODUCTION

Hemielliptic dielectric lenses are vital building blocks of dielectric-based antennas operating at millimetre (mm) and sub-mm wavelengths. Integration of such lenses with printed sensors improves matching, eliminates losses related to the surface wave excitation, and enables one to improve the aperture efficiency of such antennas [1, 2]. Moreover, it provides an opportunity to improve or reshape radiation patterns of primary feeds according to given specifications [3].

In the ray-tracing approximation, hemielliptic lenses behave similar to parabolic reflectors: they transform part of a spherical wave radiated by a point source placed in the focus into a collimated beam [4]. For large-size lenses made of low-density materials, this optical-type mechanism works rather well. This enables development of wide variety of integrated antennas for terrestrial and satellite applications, e.g. [1-3].

If the lens electrical size becomes small (several wavelengths) and/or the permittivity of the lens material becomes high ($\varepsilon > 4$) while the loss tangent is small, the influence of the internal resonances on the lens behavior becomes significant [5]. When excited, the resonances strongly affect the radiation properties of the lens antennas [6, 7]. Proper accounting for both optical and modal features of lenses at the design stage is therefore extremely important.

In the paper, we first illustrate the optical (focusing) and modal (resonant) features intrinsic to hemielliptic dielectric lenses. This is done in two-dimensional formulation using highly efficient software developed based the Muller boundary integral equations [6-8]. Then we illustrate the hazard effects of internal resonances on the radiation characteristics of lens-based antennas integrated with localized sources and outline the ways of minimizing their parasitic impact on antenna performance via optimization of the lens profile and proper adjustment of the feed parameters. Finally, we briefly discuss the computational difficulties related to accurate description of the resonances in open dielectric scatterers.

## II. LENS MODEL & METHODS OF ANALYSIS

The lens is modelled by a homogeneous dielectric cylinder whose contour is combined from two curves, namely one half-ellipse whose eccentricity equals the inverse of the material refracting index ($e = 1/\varepsilon^{1/2}$, $l_2 = [\varepsilon/(\varepsilon-1)]^{1/2}$), and one half-superellipse (rectangle with rounded corners), smoothly joint at the points (x,y) = (0, ± *a*), where *a* is the minor semi-axis of the ellipse (Fig. 1). Hereafter, these points are referred as the "edge" of the lens aperture because the focusing ability of the lens is determined by its elliptical front part whereas the extension is used to place the feed at the right (focal) distance.

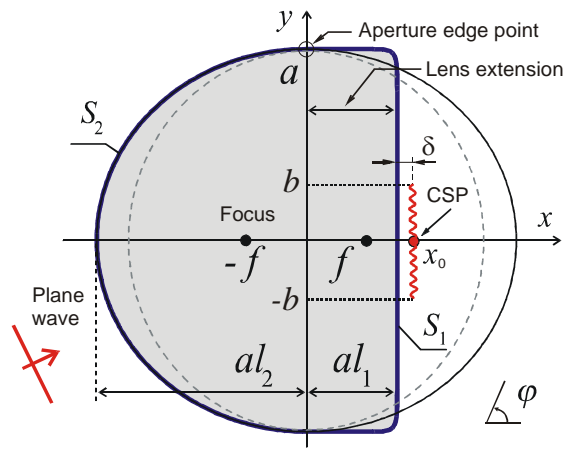

Fig. 1. Geometry and notations of the 2-D model of an extended hemielliptic dielectric lens antenna excited either by a plane wave (receiving mode) or an aperture feed simulated by a complex source point beam (emitting mode).

The lens is illuminated either by a plane wave or a beam-field produced by the so-called complex source point (CSP) feed [9-10]. The two types of incident field are used to study the focusing and collimating properties of the lens in the receiving and emitting modes, respectively.

The near- and far-field characteristics of the two-dimensional antenna model are calculated with the highly accurate in-house software developed based of the Muller boundary integral equations (MBIE) and method of analytical regularization (MAR). Details of the algorithm as well as demonstration of its efficiency in describing the true electromagnetics properties of finite-size open dielectric scatterers can be found in [6-8].

### III. FOCUSING AND COLLIMATING CHARACTERISTICS OF HEMIELLIPTIC DIELECTRIC LENSES

Once again, we remind that any dielectric lens is an open dielectric resonator capable of supporting infinite number of natural modes whose resonance frequencies and quality factors depend on the lens parameters, namely: size, shape, and material. For hemielliptic lenses the most important resonances are the so-called half-bowtie (HBT) ones [11] having specific triangular patterns detectable in the near-field zone (Fig. 2). These modes are classified by the number of in-resonance near field variations along a certain triangular contour marked by the black dashed line in Fig. 2b. As any other resonant mode an $HBT_{n,m}$ one is excited each time when the incident field frequency hits the real part of the complex-valued frequency of the corresponding natural mode. In the ray tracing approximation this corresponds to $A = n\lambda_e/2 \cup B = m\lambda_e/2$, where $\lambda_e$ is the wave-length in dielectric. This condition is periodically satisfied at different frequencies for lenses of any size and having different extensions (Fig. 3). Thus the HBT resonances are always involved in the performance of any hemielliptic lens antenna. Their quality factor and the impact on the radiation characteristics increase proportionally to the permittivity of lens material. The latter becomes noticeable already for quartz lenses ($\varepsilon = 3.8$) [12], whereas for lenses made of denser material ($\varepsilon \geq 10$) such as silicon ($\varepsilon = 11.7$) the modal (resonant) behaviour becomes dominant. In the receiving mode, it leads to a strong distortion and shift of the focal spot position (compare [13] and [5, 6]). In the emitting mode, this affects antenna matching and results in main-beam directivity degradation as well as increase of the side-lobes level [7, 12]. The latter phenomenon is illustrated in Fig. 4. As one can see, excitation of an HBT resonance in a silicon lens results in deformation of the main beam and 5 dB increase of the side-lobe level.

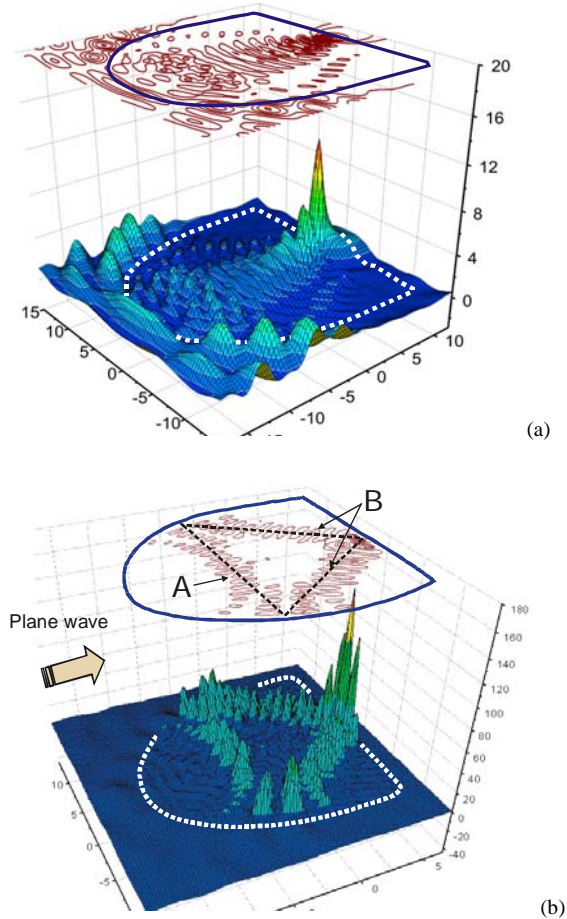

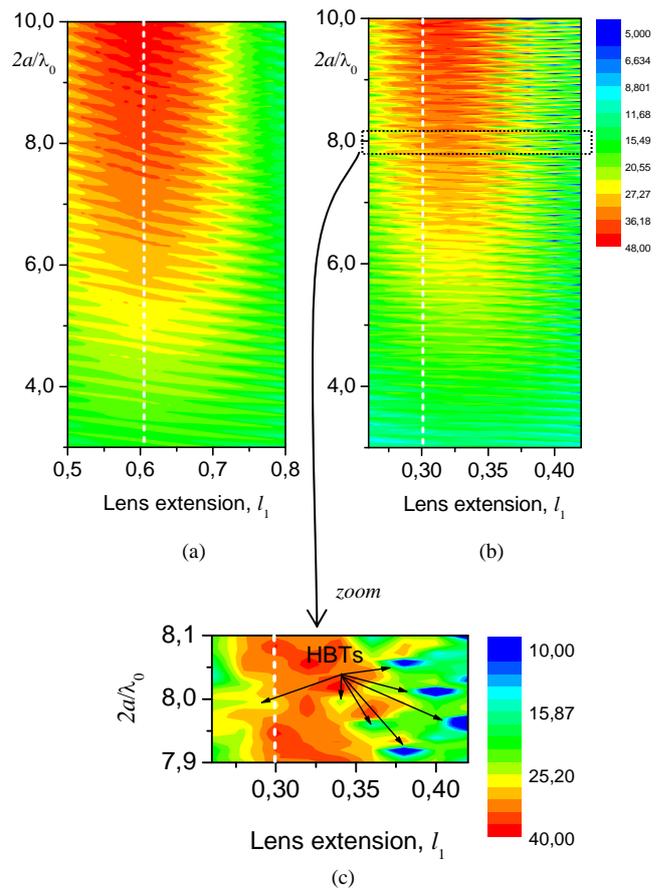

Fig. 2. Near-field intensity maps for the hemielliptic lenses illuminated by the unit-amplitude plane E-wave ($\varphi = 0°$): (a) focusing in a quartz lens, (b) HBT resonance excited within a silicon lens. The white dashed lines at the relief figures and the blue solid lines at the top contour figures, which are a projection of the relief figure to the top plane, indicate the lens contour.

Fig. 3. Color maps surface graph of the broadside directivity versus two parameters determining the lens geometry, namely lens bottom extension (horizontal axis) and width of the lens flat bottom (vertical axis): (a) quartz, (b) silicon, (c) zoom of the graph (b). Vertical white dashed line indicated the extension value corresponding to the standard cut-through-focus lens design.

## IV. HEMIELLIPTIC DLA DESIGN PECULIARITIES

As it was already mentioned, elliptical dielectric lenses have the ability to collect rays propagating parallel to their axis of symmetry into their focus, e.g. [1, 4]. Reciprocally, in emitting mode, this shape is expected to provide a locally plane wave in the radiating aperture of the hemielliptic DLA. This ray-tracing focusing/collimating rule is well satisfied for large-size lenses whose local surface curvature is negligible. In this case the Snell's laws are sufficient to describe the scattering/refraction phenomena at the boundary of the lens and its profile contour (the ellipse whose eccentricity equals the inverse of the refraction coefficient) can be derived analytically. For lenses of smaller size the elliptical shape no longer remains optimal (i.e. capable of providing the best broadside directivity). Nevertheless the small size of the lens can be compensated via shaping the lens profile [14-19]. Note that optimization procedure of such lenses is a tricky task due to the aforementioned interplay of the ray-like and resonant features that often appears to be a bottleneck for conventional techniques, e.g. [6].

Another important aspect in the design of DLAs that often escapes attention is the necessity of proper mutual adjustment of the lens size/shape and primary feed pattern. As it was recently demonstrated [12], the best performance of hemielliptic DLAs (in terms of the main-beam directivity value) is achieved when the optimal edge illumination is provided. Although this recommendation is in line with the general theory of aperture antennas, its application to design of integrated DLAs is not that straightforward. This is due to the following important differences between the reflector and integrated lens antennas. First, both the electrical size and the focal distance of hemielliptic DLAs are usually much smaller than that of reflectors [2], and thus the feed is never far away from the lens. Second, unlike a gently curved metal reflector, hemielliptic dielectric lens is an open dielectric resonator which is capable of supporting HBT modes with relatively high Q-factors. Finally, for DLAs, the focal distance and thus the favourable feed location depend on the lens material. This happens because, in the geometrical optics approximation, the eccentricity of elliptical lens is determined by its material permittivity [1, 4]. These essential distinctions between reflector antennas and DLAs result in the necessity to accurately adjust the edge illumination depending on the lens size and material [12]. As one can see in Fig. 5, for a DLA made of quartz the highest directivity is achieved when fed by primary feeds whose directivity in free space equals $D_0 = 5 \sim 6$ dB, whereas for a silicon DLA of the same size best directivity is provided if it is excited by an almost omnidirectional feed ($D_0 = 2 \sim 3$ dB). It is also clearly seen that the optimal edge illumination does not prevent from excitation of HBT resonances although it reduces their impact by requesting a decay of field intensity at the lens aperture edges comparing to the origin [12]. Another natural mechanism of reduction of the HBT resonances impact is the presence of losses in the lens material (Fig. 6). Additional suppression can be achieved by introducing a matching layer either in the form of dielectric films [20-22] or surface corrugations [23].

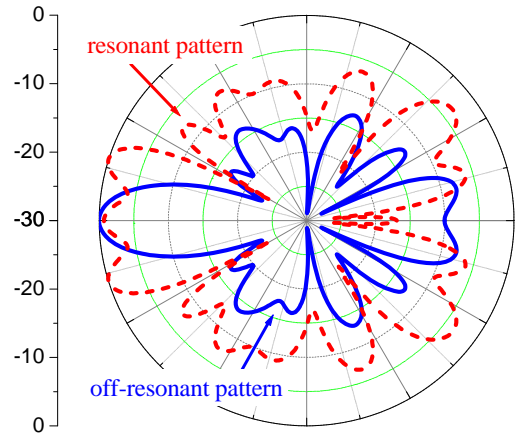

Fig. 4. Normalized far-field radiation patterns of hemielliptic silicon DLAs excited by unidirectional line current feeds placed close to lens bottom ($\delta = \lambda_0/10$). The two curves correspond to antennas with lenses having equal bottom size ($\varnothing = 3\lambda_0$) and slightly different bottom extensions: (solid line) cut-through-focus design, $l_1=0.31$, (dashed line) resonant lens supporting an HBT resonance, $l_1=0.36$.

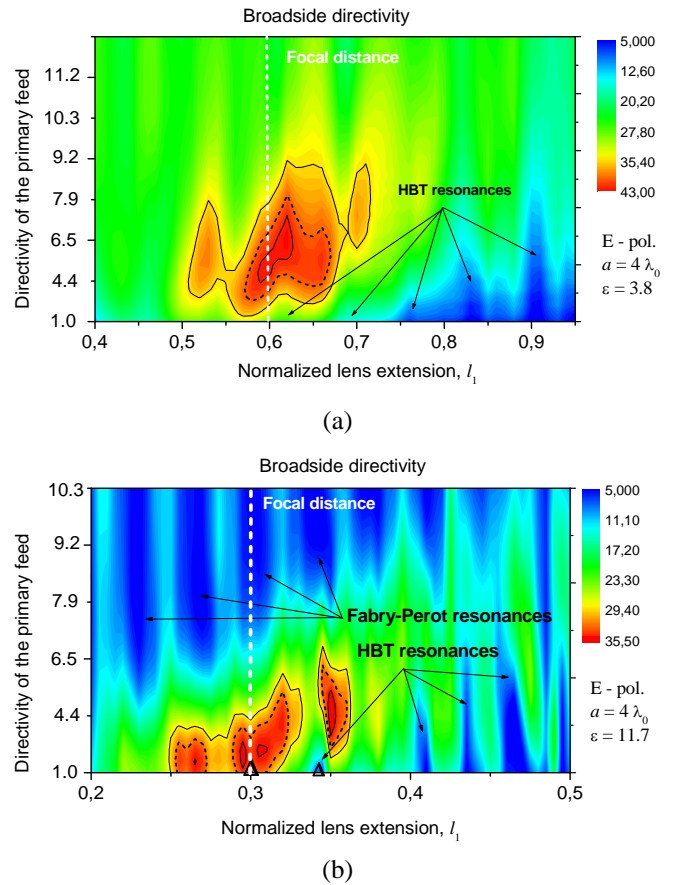

Fig. 5. Broadside directivity of the silicon hemielliptic DLA ($ka = 25.12$) vs. normalized lens extension ($l_1$) and directivity of the primary feed, simulated by CSP beam. For clarity, only top 2%, 10%, and 20% value contours are shown. The white dashed lines indicate the length of the lens extension corresponding to the cut-through-focus design.

Until now we discussed performance of the *hemielliptic* DLAs which are the most popular solution for various practical applications addressing the point-to-point communication or scanning scenarios. For the integrity of the discussion it is also useful to mention the so-called shaped DLAs having lenses with profiles specially designed to satisfy some specific requirements such as shaped-beam formation [3, 15, 24] and/or improved angular characteristics [18, 25]. Design of such antennas usually involves optimization of the lens shape as well as topology and parameters of the feeding structure. This multi-parameter problem can be successfully handled using local and/or global optimization techniques [14-18, 25, 26].

There are two critical points in the design procedure that are worth to be mentioned in the scope of the current paper. First, as any optimization problem DLA design is based on iterative solution of the direct diffraction problem. Thus there is always a compromise between the accuracy of the diffraction problem solution (controlled by the number and size of ray tubes used in GO/PO techniques [14, 15], mesh size for MoM and FDTD approaches, e.g. [7, 19], or the matrix truncation number for MBIE-based algorithms [6-8]) and the computational time. This trade-off is often solved in the favour of the computation time reduction that may cause a fail of the design procedure. This is because arbitrary shaped-lenses (especially those having small size and made of dense material) are capable of supporting modes whose resonant frequencies and quality factors cannot be predicted in advance, (except by solving the relevant eigen-value problems that is an independent and complicated electromagnetic problem). Excitation of strong internal resonances may lead to uncontrollable errors entering the solution that misguides the optimization procedure by over- or under-estimating the quality of the considered design. There is no simple solution to this problem because it comes from the nature of each electromagnetic solver. For instance, it is possible to compensate the inaccuracies of the geometrical/physical optical techniques via accounting for multiple reflections, introducing corrections in the regions where the total reflection appears [27-29], etc., but there is no remedy for the case when radius of curvature of the lens surface becomes comparable to the wavelength. Even full-wave methods such as FDTD and MoM may fail to describe accurately high-$Q$ resonances in dielectric scatterers due to staircase errors [7, 30, 31]. Therefore it is extremely important (a) to pay the appropriate attention to inherent resonant properties of open dielectric resonators, and (b) to develop and apply numerical algorithms with in-built criteria of accuracy (i.e. capable to estimate the computational error for the given set of the problem parameters). Our recommendation to this end is to use the algorithms developed based on the MBIE / MAR approach, whose guaranteed monotonic convergence enables one to control the accuracy of the numerical solution for any set of the DLA parameters.

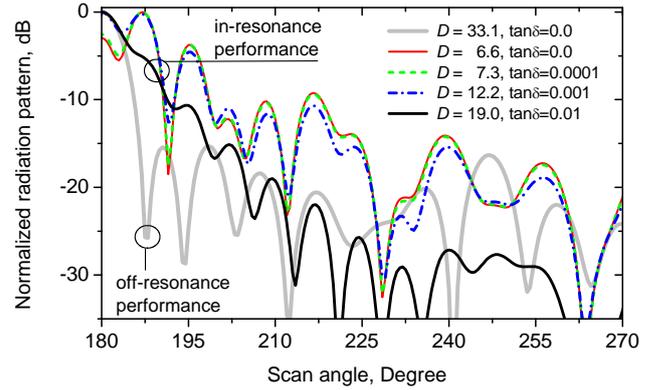

Fig. 6. Normalized far-field radiation patterns of hemielliptic silicon DLAs whose configurations are depicted in Fig. 5b by triangles. Thick grey line corresponds DLA whose extension is chosen to provide the highest broadside directivity (marked as "off-resonance"). The family of curves marked as "in-resonance" corresponds to the design supporting an HBT resonance; the curves differ by the value of losses in the material defined by tanδ.

## V. Conclusions

The optical and modal features of hemielliptic lenses, typically used as building blocks of integrated DLAs operating at the mm and sub-mm wavelength ranges, have been carefully studied using highly-accurate in-house software developed based on the MBIE / MAR approach. It was demonstrated that HBT resonances (that can be easily identified by their specific triangular near-field patterns) are inherently involved in the performance of hemielliptic DLAs. Their impacts on the DLA performance and difficulties related to their accurate characterisation have been discussed on examples of hemielliptic quartz and silicon lenses.

More details on the subject will become available in [32].

## Acknowledgment

This work was supported in part by Université Européenne de Bretagne, Rennes, France, by the Ministry of Science and Education of Ukraine via project M/146-2009 and by the North Atlantic Treaty Organization via grant RIG983313.

## References

[1] D.F. Filipovic, S.S. Gearhart, and G.M. Rebeiz, "Double slot antennas on extended hemispherical and elliptical silicon dielectric lenses," *IEEE Trans. Microw. Theory Tech.*, vol. 41, no. 10, pp. 1738–1749, Oct. 1993.

[2] G. Godi, R. Sauleau, and D. Thouroude, "Performance of reduced size substrate lens antennas for millimeter-wave communications," *IEEE Trans. Antennas Propag.*, vol. 53, no. 4, pp. 1278–1286, Apr. 2005.

[3] C.A. Fernandes, "Shaped-beam antennas," in *Handbook of Antennas in Wireless Communications*, L.C. Godara, Ed., New York: CRC Press, 2002, ch. 15.

[4] P. Varga, "Focusing of electromagnetic radiation by hyperboloidal and ellipsoidal lenses," *J. Opt. Soc. Am. A*, vol. 19, pp. 1658–1667, 2002.

[5] A.V. Boriskin, A.I. Nosich, S.V. Boriskina, T.B. Benson, P. Sewell, and A. Altintas, "Lens or resonator? - electromagnetic behavior of an extended hemielliptical lens for a sub-mm wave receiver," *Microw. Optical Tech. Lett.*, vol. 43, no. 6, pp. 515–518, Jun. 2004.

[6] A.V. Boriskin, G. Godi, R. Sauleau, and A.I. Nosich, "Small hemielliptic dielectric lens antenna analysis in 2-D: boundary integral equations versus geometrical and physical optics," *IEEE Trans. Antennas Propag.*, vol. 56, no. 2, pp. 485–492, Feb. 2008.

[7] A.V. Boriskin, A. Rolland, R. Sauleau, and A.I. Nosich, "Assessment of FDTD accuracy in the compact hemielliptic dielectric lens antenna analysis," *IEEE Trans. Antennas Propag.*, vol. 56, no. 3, pp. 758–764, Mar. 2008.

[8] S.V. Boriskina, T.M. Benson, P. Sewell, and A.I. Nosich, "Accurate simulation of 2D optical microcavities with uniquely solvable boundary integral equations and trigonometric-Galerkin discretization," *J. Opt .Soc. Am. A*, vol. 21, no. 3, pp. 393–402, 2004.

[9] P.D. Einziger, Y. Haramaty, and L.B. Felsen, "Complex rays for radiation from discretized aperture distributions," *IEEE Trans. Antennas Propag.*, vol. 35, no. 9, pp. 1031–1044, Sep. 1987.

[10] E. Heyman and L.B. Felson, "Gaussian beam and pulsed-beam dynamics: complex-source and complex-spectrum formulations within and beyond paraxial asymptotics," *J. Opt. Soc. Am. A*, vol. 18, no. 7, pp. 1588–1611, 2001.

[11] J. Wiersig, "Formation of long-lived, scarlike modes near avoided resonance crossings in optical microcavities," *Phys. Rev. Lett.*, vol. 97, 253901, 2006.

[12] A.V. Boriskin, R. Sauleau, and A.I. Nosich, "Performance of hemielliptic dielectric lens antennas with optimal edge illumination," *IEEE Trans. Antennas Propag.*, vol. 57, no. 7, pp. 2193–2198, Jul. 2009.

[13] A.V. Boriskin, R. Sauleau, and A.I. Nosich, "Exact off-resonance near fields of small-size extended hemielliptic 2-D lenses illuminated by plane waves," *J. Opt. Soc. Am. A*, vol. 26, no. 2, pp. 259–264, Feb. 2009.

[14] B. Barès, R. Sauleau, L. Le Coq, and K. Mahdjoubi, "A new accurate design method for millimeter-wave homogeneous dielectric substrate lens antennas of arbitrary shape," *IEEE Trans. Antennas Propag.*, vol. 53, no. 3, pp. 1069–1082, Mar. 2005.

[15] R. Sauleau and B. Barès, "A complete procedure for the design and optimization of arbitrary shaped integrated lens antennas," *IEEE Trans. Antennas Propag.*, vol. 54, no. 4, pp. 1122–1133, Apr. 2006.

[16] A.V. Boriskin and R. Sauleau, "Dielectric lens antenna size reduction due to the shape optimization with genetic algorithm and Muller boundary integral equations," *Proc. European Microwave Conference (EuMC-06)*, Manchester (UK), Sep. 2006, pp. 287–290.

[17] G. Godi, R. Sauleau, L. Le Coq, and D. Thouroude, "Design and optimization of three-dimensional integrated lens antennas with genetic algorithm," *IEEE Trans. Antennas Propag.*, vol. 55, no. 3, pp. 770–775, Mar. 2007.

[18] A.V. Boriskin and R. Sauleau, "Synthesis of arbitrary-shaped lens antennas for beam-switching applications," *Proc. European Microwave Conference (EuMC)*, Paris (France), Sept. 26-30, 2010, paper ID#EuMC45-3.

[19] A. Rolland, R. Sauleau, and M. Drissi, "Design of H-plane shaped flat lenses using a 2-D approach based on FDTD and genetic algorithm," *Proc. European Conf. on Antennas and Propagation (EuCAP-10)*, Barcelona (Spain), Apr. 12-16, 2010.

[20] D.F. Filipovic, G.P. Gauthier, S. Raman, and G.M. Rebeiz, "Off-axis properties of silicon and quartz dielectric lens antennas," *IEEE Trans. Antennas Propag.*, vol. 45, no. 5, pp. 760–767, May 1997.

[21] N.T. Nguyen, R. Sauleau, and C.J.M. Perez, "Very broadband extended hemispherical lenses: role of matching layers for bandwidth enlargement," *IEEE Trans. Antennas Propag.*, vol. 57, no. 7, pp. 1907–1913, Jul. 2009.

[22] A. Neto, "UWB, non dispersive radiation from the planarly fed leaky lens antenna. Part 1: theory and design," *IEEE Trans. Antennas Propag.*, vol. 58, no. 7, pp. 2238–2247, 2010.

[23] N.T. Nguyen, R. Sauleau, N. Delhote, D. Baillargeat, and L. Le Coq, "Design and characterization of 60-GHz integrated lens antennas fabricated through ceramic stereolithography," *IEEE Trans. Antennas Propag.*, vol. 58, no. 8, pp. 2757–2762, 2010.

[24] R.I. Henderson, "Millimetre-wave reflectarray fed by a diffraction-shaped dielectric lens," *Proc. European Conf. Antennas Propag. (EuCAP-09)*, Berlin (Germany), Mar. 23-27, 2009, pp. 2458–2462.

[25] H. Kawahara, H. Deguchi, M. Tsuji, and H. Shigesawa, "Design of rotational dielectric dome with linear array feed for wide-angle multibeam antenna applications," *Electronics and Communications in Japan*, part 2, vol. 90, no. 5, pp. 49–57, 2007.

[26] E. Lima, J.R. Costa, M.G. Silveirinha, and C.A. Fernandes, "ILASH - Software tool for the design of integrated lens antennas," *Proc. IEEE Antennas Propag. Symp (APS-08)*, 5-11 July 2008, pp. 1–4.

[27] D. Lemaire, C. A. Fernandes, P. Sobieski, and A. Barbosa, "A method to overcome the limitations of GO in axis-symmetric dielectric lens shaping," *Int. J. Infrared and Millimeter Waves*, vol. 17, no. 8, pp. 1377–1390, 1996.

[28] D. Pasqualini and S. Maci, "High-frequency analysis of integrated dielectric lens antennas," *IEEE Trans. Antennas Propag.*, vol. 52, no. 3, pp. 840–847, Mar. 2004.

[29] A.P. Pavasic, D.L. del Rio, J.R. Mosig, and G.V. Eleftheriades, "Three-dimensional ray-tracing to model internal reflections in off-axis lens antennas," *IEEE Trans. Antennas Propag.*, vol. 54, no. 2, pp. 604–613, Feb. 2006.

[30] A.V. Boriskin, A. Rolland, R. Sauleau, and A.I. Nosich, "Test of the FDTD accuracy in the analysis of the scattering resonances associated with high-Q whispering-gallery modes of a circular cylinder," *J. Opt. Soc. Am. A*, vol. 25, no. 5, pp. 1169–1173, May 2008.

[31] G.L. Hower, R.G. Olsen, J.D. Earls, and J.B. Schneider, "Inaccuracies in numerical calculations of scattering near natural frequencies of penetrable objects", *IEEE Trans. Antennas Propag.*, vol. 41, no. 7, pp. 982–986, Jul. 1993.

[32] A.V. Boriskin and R. Sauleau, "Drastic influence of the half-bowtie resonances on the focusing and collimating capabilities of 2-D extended hemielliptical and hemispherical dielectric lenses," *J. Opt. Soc. Am. A*, vol. 27, no. 11, 2010.